# Formation of copper boride on Cu(111)

Chengguang Yue,[1†] Xiao-Ji Weng,[2†] Guoying Gao,[1] Artem R. Oganov,[3] Xiao Dong,[2] Xi Shao,[1] Xiaomeng Wang,[4] Jian Sun,[4] Bo Xu,[1] Hui-Tian Wang,[2,4] Xiang-Feng Zhou,[1,2*] and Yongjun Tian[1]

[1]*Center for High Pressure Science, State Key Laboratory of Metastable Materials Science and Technology, School of Science, Yanshan University, Qinhuangdao 066004, China*

[2]*Key Laboratory of Weak-Light Nonlinear Photonics, School of Physics, Nankai University, Tianjin 300071, China*

[3]*Skolkovo Institute of Science and Technology, Skolkovo Innovation Center, 3 Nobel Street, Moscow 121025, Russia*

[4]*National Laboratory of Solid State Microstructures, Collaborative Innovation Center of Advanced Microstructures, Nanjing University, Nanjing 210093, China*

[†]These authors contributed equally to this work

## Abstract

Boron forms compounds with nearly all metals, with notable exception of copper and other group IB and IIB elements. Here, we report an unexpected discovery of ordered copper boride grown epitaxially on Cu(111) under ultrahigh vacuum. Scanning tunneling microscopy experiments combined with *ab initio* evolutionary structure prediction reveal a remarkably complex structure of 2D-$Cu_8B_{14}$. Strong intra-layer p–d hybridization and a large amount of charge transfer between Cu and B atoms are the key factors for the emergence of copper boride. This makes the discovered material unique and opens up the possibility of synthesizing ordered low-dimensional structures in similar immiscible systems.

[†]These authors contributed equally to this work
*Corresponding author.
xfzhou@nankai.edu.cn or zxf888@163.com

# 1. Introduction

Metal borides attract much interest because of their unique properties, including superconductivity, superhardness, high melting points, chemical inertness, and quantum topological properties [1-7]. Although more than 1200 binary borides are included in the Inorganic Crystal Structure Database [8], borides of group IB and IIB metals are practically unknown, due to small difference in electronegativities and large difference in atomic sizes of boron and these metals. There are only a few reports of boron-doped copper or copper-doped boron, without reliable stoichiometry, e.g., $CuB_{\sim22}$, $CuB_{\sim23}$, or $CuB_{\sim28}$ [9-12]. It is well known that boron is a metalloid element with electronegativity very close to that of copper (Pauling electronegativity of 2.04 and 1.9, respectively), implying little charge transfer [13]. Ordered alloys and normal covalent bonding between Cu and B would also be difficult to form because of the large atomic size mismatch of $\Delta V / \bar{V} = 129\%$, leading to the positive energies of formation at ambient conditions [8, 14]. Moreover, calculations based on density functional theory (DFT) showed that copper borides are also unstable at pressures at least up to 30 GPa [14]. Perhaps such compounds can be synthesized at higher pressures, or in a low-dimensional form (e.g. on a specific substrate). Such a strategy worked very well in the case of sodium chlorides because three-dimensional (3D) $NaCl_3$ and $Na_3Cl$ phases were experimentally realized at pressures above 60 GPa [15], while two-dimensional (2D) $Na_2Cl$ and $Na_3Cl$ were successfully synthesized on a graphene substrate in dilute salt solution [16]. Although stable copper-rich borides at high pressure has never been reported experimentally, they may be formed at at extreme conditions, i.e., above 100

GPa. The most recent prediction of 2D copper borides revealed its feasibility at the scale of atomic thickness and some striking properties [17]. Here we conducted independent experiments by depositing boron on the Cu(111) surface and obtained samples with an unexpected herringbone-like structure. Part of experimental results are similar to the published work by Wu *et al*, what they declared the successful synthesis of borophene on Cu(111) [18]. However, as we show below, scanning tunneling microscopy (STM) measurements and DFT simulations elucidated this structure as a 2D copper boride.

## 2. Methods

The experiment was performed in an ultrahigh vacuum system with conjoint molecular beam epitaxy (MBE) chamber and sample characterization chamber. A single crystal Cu(111) substrate (surface roughness < 10 nm and orientation accuracy < 0.1°, MaTecK GmbH) was treated with multicycles of $Ar^+$ sputtering and annealing to achieve a clean and smooth surface, and then transferred into the MBE chamber (with a base pressure of $1\times10^{-10}$ mbar) for boron deposition. Pure boron was evaporated with a high temperature effusion cell (CreaTec GmbH) at the temperature higher than 2020 K. Two different procedures were used to prepare copper boride: In the first one, boron was deposited on the substrate that was kept at room temperature, with subsequent annealing at 500 and 540 K; in the other one, boron was deposited on the substrate kept at 740 K. In the latter case, Cu(111) surface was fully covered by copper boride, without annealing, after 25 minutes of B deposition. The prepared samples were cooled to room temperature and transferred into the characterization chamber for further study. The characterization chamber with a base pressure of 5×10$^{-11}$ mbar is equipped with an Aarhus VT-STM and SPECS ErLEED. At each stage of sample preparation, the surface (such as clean Cu(111), room temperature deposited surface with or without annealing, 740 K deposited surface) was studied with STM and LEED/AES. All reported STM images were collected at room temperature with a constant current mode unless otherwise specified. The tunneling bias is given for the sample with the STM tip virtually grounded. AES spectra were taken with an incident electron energy of 1000 eV.

Surface structure searches were performed using the *ab initio* evolutionary algorithm USPEX [19-22]. The initial structures were produced with assigned plane group symmetries (*p2mg*) on the modified $2\times\sqrt{37}$ Cu(111) substrate. The stopping criterion for evolutionary searches was set as 30, which means that the evolutionary searches will be stopped if the best structure (total energy) did not change for 30 generations (each generation contains 60 structures) or when the total number of generations have been expired (50 generations). The relaxed energies were adopted as a criterion for parent structures selection to generate new structures by various evolutionary operators, such as random structure generator, heredity and mutations. The user-defined initial thicknesses of surface and buffer regions were set to 3 Å and 1 Å, respectively, and allowed to change during the relaxation. Structure relaxations and electronic calculations were performed using the projector-augmented wave [23] method as implemented in the VASP package [24]. The exchange-correlation energy was treated within the generalized gradient approximation (GGA), using the functional of Perdew,

Burke, and Ernzehof (PBE) [25]. For structure search, the plane-wave cutoff energy of 450 eV, uniform Γ-centered k-point grids with the resolution of $2\pi \times 0.04$ Å$^{-1}$, the convergence criterion of $10^{-4}$ eV for terminating the electronic self-consistency cycle (SCC) were used. Further accurate energies and electronic properties were applied for the most stable structures with the energy cutoff of 500 eV, k-point resolution of $2\pi \times 0.03$ Å$^{-1}$, force convergence criterion of $10^{-2}$ eV/Å, and SCC energy convergence criterion of $10^{-6}$ eV. The phonon dispersion curves were computed using the supercell method (3×2×1) implemented in the PHONOPY package [26].

## 3. Results and Discussion

First, we deposited boron on Cu(111) substrate at room temperature. After depositing for 10 minutes, STM showed only disordered phases on the surface (Fig. S1a). By annealing the samples at the temperature of ~500 K, an ordered phase (copper boride) was formed, bringing about a significant surface reconstruction, and both the ordered and disordered phases coexisted on the surface with sawtooth-shaped step edges (Figs. S1b and S1c). As we increased the annealing temperature to ~540 K, the disordered phase transformed completely into the ordered one (Fig. 1a). High-resolution STM image revealed that copper boride can be grown as a large-area layer on Cu(111), and displays a modulated wavy pattern (Fig. 1b). The step height between adjacent clean terraces is 0.21 nm, which consistents with the interplanar spacing of Cu(111) planes, verifying the accuracy of the measurements (Fig. 1c). Furthermore, the height difference between the top (bottom) layer of copper boride and Cu(111) terraces is ~0.14 (0.07) nm (Fig. 1c), indicating that copper boride with a monolayer structure has a weaker tunneling current than Cu(111). Large area of ordered copper boride can also be synthesized by directly depositing boron on Cu(111) at ~740 K without annealing (Fig. 1d and Fig. S1d). The results of the high-temperature growth and room-temperature growth show differences in the size and distribution of disordered regions (Fig. 1d and Fig. S1). Nevertheless, all the ordered structures are identical despite different growth approaches (Figs. 1b and 1e). Figure 1f shows Auger electron spectroscopy (AES) spectra taken from the surface before and after boron deposition. Before deposition, only signals of super Coster–Kronig transition of Cu ($M_1M_{45}M_{45}$ and $M_{23}M_{45}M_{45}$ peaks at 61.0 eV and 106.0 eV, respectively) were detected without any peaks from oxygen, carbon or other contamination. After boron deposition, these peaks are detected at 60.8 eV and 105.7 eV respectively (Fig. S2), showing the likelihood variation of chemical state of copper. Furthermore, additional peaks near 180 eV appear after boron deposition, corresponding to B KLL peak [27, 28]. A closer examination reveals split boron peaks (at 178.2 and 182.3 eV, inset of Fig. 1f), which was absent in Ag(111)-borophene system [29, 30]. Similar peak splitting was also observed in C KLL of nickel carbide [31]. The splitting of B KLL peaks in our sample originates from the interaction between B and Cu, maybe a hint on the existence of copper boride on Cu(111) surface.

High-resolution STM images show distinct atomic zigzag patterns with lattice constants of $a$ = 0.55 nm , $b$ = 1.50 nm and $\alpha$ = 90°, or another equivalent cell with parameters of $a'$ = 0.55 nm, $b'$ = 1.59 nm and $\alpha'$ = 70° (Fig. 2a). The measured angles between $a'$ axis and [1$\bar{1}$0] direction, or between $b'$ axis and [11$\bar{2}$] direction of Cu(111), are ~5.9° or ~13.3°, respectively. Hence the lattice constants of a 4×1 supercell of copper boride match $\sqrt{73}\times\sqrt{39}$ Cu(111) very well with the corresponding lattice constants of $4a' = 2.184$ nm, $b' = 1.596$ nm, $\alpha' = 70.28°$ (Fig. 2a). In addition, low energy electron diffraction (LEED) patterns show that the strongest diffraction spots (circled in blue in Fig. 2c) come from the 1×1 Cu(111) substrate, whereas other spots are related to unkown structures and some of these spots are shown more clearly in the LEED image with lower energy (Fig. S3). Regardless of minor distortion from the measured image, the whole patterns indicate six-fold domains from the overlay structure (labeled by the yellow and red ovals in Fig. 2c). For comparison, using the refined lattice constants $a = 0.546$ nm, $b = 1.503$ nm and $\alpha = 90.28°$, we simulated the LEED patterns from Cu(111) overlapping with copper boride [32]. Note that six types of copper boride domains are included in the simulation: two mirror-symmetric ones for each of three equivalent high-symmetry directions on the Cu(111) surface. The results shown in Fig. 2d are in good agreement with the experimental patterns (Fig. 2c), proving the lattice constants derived from STM measurement is unambiguous.

With precise calibration of lattice vectors, the next challenge was to determine the accurate composition and atomic positions in copper boride. The high-resolution STM image (Fig. 2a) indicates eight bright spots per unit cell (where one building block is highlighted by solid blue circles). These spots are distributed approximately uniformly and assembled into two nearly vertical line segments with length of ~1 nm, forming an extended structure. The distance of ~0.25 nm between two adjacent spots is very close to the shortest Cu–Cu distance (~0.256 nm) in metallic copper. To clarify the structure of copper boride and in particular, the origin of these bright spots (and to check whether this pattern could correspond to an intrinsic reconstruction of the Cu(111) surface), we performed first-principles calculations. According to the positions indicated by STM image, an additional Cu monolayer featuring zigzag chains of Cu atoms was superimposed on the $\sqrt{73}\times\sqrt{39}$ Cu(111) substrate. This special configuration is mechanically unstable and transforms into disordered nano-fragments (Fig. S4). Most probably, these bright spots should be attributed to Cu atoms, but with stabilization due to Cu-B interactions. The atomic zigzag with eight Cu atoms per period, together with the measured AES data, imply the formation of a 2D copper boride with stoichiometry $Cu_8B_x$ ($x$ denotes the number of B atoms per unit cell).

The first estimate of stoichiometry can be made using planar atomic density. The $v_{1/6}$-borophene ($v_{1/6}$-$B_5$) was predicted to be the most stable borophene structure on Cu(111) [33], and atomic densities for $v_{1/6}$-borophene and 1×1 Cu(111) are 0.03 nm$^2$/atom and 0.0567 nm$^2$/atom. Given lattice constants of 2D-$Cu_8B_x$ ($a$ = 0.546 nm, $b$ = 1.503 nm and α = 90.28°), the estimated value of $x$ is about 12.2. To get the exact number and positions of boron atoms, we use the evolutionary structure prediction. Full structure prediction for the 4×1 supercell of $Cu_8B_x$ on $\sqrt{73}\times\sqrt{39}$ Cu(111) is impractical: the whole system would contain more than 250 atoms, assuming a three-layered substrate. After carefully comparing the lattice constants of 2D-$Cu_8B_x$ and Cu(111) substrate, we found that the lattice constants of $2\times\sqrt{37}$ Cu(111) ($a''$ = 0.511 nm, $b''$ = 1.555 nm and α = 94.72°) are closest to those of 2D-$Cu_8B_x$ (Fig. S5), and performing a structure prediction for this system is feasible. Hence *ab initio* evolutionary algorithm USPEX was employed to do surface structure searches for the likeliest stoichiometries: $Cu_8B_{10}$, $Cu_8B_{12}$, $Cu_8B_{14}$, and $Cu_8B_{16}$ on a distorted $2\times\sqrt{37}$ Cu(111) substrate (Fig. S5). Among the lowest-energy structures for these compositions, only 4×1 $Cu_8B_{14}$ on Cu(111) is consistent with experimental results (Fig. 2a and 2b).

2D-$Cu_8B_{14}$ has a plane group *p2mg* symmetry with the fixed lattice constants of *a* = 0.546 nm and *b* = 1.503 nm (Fig. S6). Three nonequivalent Cu sites are Cu1 (0.0, 0.0), Cu2 (0.6473, 0.1182), and Cu3 (0.3542, 0.25), four nonequivalent B sites are B1 (0.3698, 0.0245), B2 (0.2177, 0.1173), B3 (0.9916, 0.1915), and B4 (0.7387, 0.25). After structure relaxation on $\sqrt{73}\times\sqrt{39}$ Cu(111), $Cu_8B_{14}$ shows a little structural corrugation, but retains its zigzag pattern. Figure 3 shows the experimental STM images obtained at different scanning conditions. As mentioned above, the bright line segments are attributed mostly to Cu atoms, while the dark line segments represent B atoms. Owing to the interactions between $Cu_8B_{14}$ and substrate, the zigzag Cu chains are slightly higher than the neighboring B chains. This leads to all B atoms being invisible except for B4 (Fig. 3a). Meanwhile, each bright corner of a zigzag chain is composed of two Cu2, one Cu3 and one B4 atoms. They are bonded together, modulated by the substrate and have a relatively higher position (Fig. 3b), which results in the brightest protrusions in the corners of wavy patterns (Fig. 3a and 3b). Depending on the tunneling conditions, such wavy patterns can partially be broken and evolve into a more complex morphology (Fig. 3c). The simulated STM images on the basis of this 2D-$Cu_8B_{14}$ are in good agreement with the experimental results (Figs. 3d, 3e and 3f), confirming the structure obtained from *ab initio* evolutionary searches. Additional comparisons of experimental results and simulated STM images shown in supplementary materials (Fig. S7) give further support to this structure. STM measurement was also performed at low temperature of about 100 K by liquid nitrogen cooling. The result shows similar modulated wavy pattern and the same lattice constants of $Cu_8B_{14}$ compared with those at room temperature (Fig. S8).

Relaxation of the freestanding 2D-$Cu_8B_{14}$ retains its *p2mg* symmetry with the lattice constants $a = 0.535$ nm and $b = 1.485$ nm, which has some lattice mismatch with the substrate. Phonon calculations show that this freestanding structure is dynamically unstable (Fig. S9), exhibiting an instability against long-wave transverse acoustic vibrations [29]. Interaction with the substrate and a minor puckering (corresponding to the out-of-plane acoustic modes) may eliminate the instability. Indeed, anisotropic corrugations are observed in the STM images with such spots in strongly contrasted brightness. These imaginary frequencies disappear when lattice constants are strained to match the experimental values (Fig. S9). Moreover, we also calculated the exfoliation energy of $Cu_8B_{14}$ on Cu(111). As shown in Fig. S9, the value of 1.81 $J/m^2$ is lower than the exfoliation energy criterion of 2.08 $J/m^2$, indicating it is potentially exfoliable [34].

The projected density of states (PDOS) clearly show that $Cu_8B_{14}$ on Cu(111) is metallic with a pronounced pseudogap and a distinct p-d hybridization in the vicinity of $E_F$. Scanning tunneling spectroscopy (the dI/dV spectrum) confirms the pseudogap around $E_F$ and metallic characteristics of $Cu_8B_{14}$, which is qualitatively consistent with the calculated PDOS (Fig. S10). The band structures of the freestanding $Cu_8B_{14}$ ignoring spin orbit coupling effect show metallicity and the existence of Dirac nodal lines near E$_F$ around the *Z* point. These nodal lines are protected by the *M$_y$* mirror symmetry and come mostly from B-*p* orbitals (Fig. 4a and 4c), which may result in superior electronic transport. Moreover, as shown in Fig. 4b, the small dispersion of the bands above the Fermi level results in a pronounced peak in the density of states, which indicates that a charge doping on this system may tune the transport properties with possible superconductivity.

For chemical bonds of Cu$_8$B$_{14}$, the B–B and Cu–B bond lengths are in the range of 0.16–0.176 nm and 0.204–0.235 nm (Fig. S6), respectively. The character of B-B and Cu-B bonds can be analyzed using electron localization function (ELF) [35]. ELF can take values between 0 and 1, and ELF value of ~0.9 represents strong covalent B-B bonds, while ELF ~0.5 shows that Cu-B bonds are delocalized (Fig. 4d). For Cu$_8$B$_{14}$ with substrate, Bader charges [36] show that charges of Cu1, Cu2, Cu3, B1, B2, B3, and B4 are ~0.26*e*, 0.29*e*, 0.25*e*, -0.28*e*, -0.19*e*, -0.22*e*, and -0.20*e*, in contrast to the average charge of ~0.06*e*/atom from the topmost Cu(111) layer. The contribution to the charge transfer from the intralayer Cu and substrate Cu atoms are 71% and 29%. While for freestanding Cu$_8$B$_{14}$, the charges of Cu1, Cu2, Cu3, B1, B2, B3, and B4 are ~0.30*e*, 0.34*e*, 0.28*e*, -0.19*e*, -0.21*e*, -0.14*e*, -0.18*e*, respectively. Consequently, the stronger intralayer p–d hybridization (formation of Cu-B bonds) and a amount of charge transfer between Cu and B atoms are crucial to the emergence of copper boride.

To further explore the thermodynamic stability of Cu$_8$B$_{14}$ on Cu(111), we calculated the energy of formation ($E_f^{Cu_{1-x}B_x}$) as

$$E_f^{Cu_{1-x}B_x} = \Delta E(Cu_{1-x}B_x) - (1-x)\Delta E(Cu) - x\Delta E(B) \tag{1}$$

where $\Delta E(Cu_{1-x}B_x)$, $\Delta E(Cu)$, and $\Delta E(B)$ is energy difference (meV/atom) between the whole system and the reference clean substrate. Cu and B represent the reference structure of Cu(111) monolayer and $v_{1/6}$-B$_5$ borophene, respectively. On the basis of Eq. (1), we constructed the convex hull by considering Cu$_8$B$_{10}$, Cu$_8$B$_{12}$, Cu$_8$B$_{14}$, Cu$_8$B$_{16}$, various borophenes of $v_{1/6}$-B$_5$, α-sheet [37], g$_{2/15}$-B$_{13}$ [38] and B$_{52}$ on Cu(111) [18] (Fig. 4e). The results show that the bare copper surface, $v_{1/6}$-B$_5$ and Cu$_8$B$_{14}$ are thermodynamically stable states, which is in accordance with our experimental results and early prediction [33].

Furthermore, to explore the chemical inertness of Cu$_8$B$_{14}$, we performed a controllable oxidation on the samples. Oxygen, with a pressure as high as $3\times10^{-6}$ mbar, was introduced into the loadlock chamber (with a base pressure of $5\times10^{-9}$ mbar) where the sample was stored. The ordered structure of Cu$_8$B$_{14}$ was retained after ten minutes oxygen exposure (Fig. S11), indicating stability of Cu$_8$B$_{14}$ against oxidation to some extent.

## 4. Conclusion

In summary, the discovery of copper boride shows that compounds that cannot exist in the bulk form (e.g., borides of group IB and IIB metals) can be formed on surfaces due to the enhanced reactivity of the surface atoms. The next challenges will be to see how the electronic performance can be tuned by chemical functionalization, to transfer the copper borides from UHV or grow different Cu−B monolayers. Creation of unusual surface compounds adds a new direction in the field of low-dimensional materials.

**Declaration of Competing Interest**

The authors declare no competing interests to this work.

**Acknowledgments** This work was supported by the National Science Foundation of China (Grants 52025026, 11674176, 11874224, and 51525205), the Tianjin Science Foundation for Distinguished Young Scholars (Grant No. 17JCJQJC44400). A.R.O. thanks Russian Science Foundation (Grant No. 19-72-30043). The authors are grateful for the technical support for Nano-X from Suzhou Institute of Nano-Tech and Nano-Bionics, Chinese Academy of Sciences (SINANO). X.D. and X.F.Z thank the computing resources of Tianhe II and the support of Chinese National Supercomputer Center in Guangzhou.

**Supplementary Information** is linked to the online version of the paper at

**Figure Legends:**

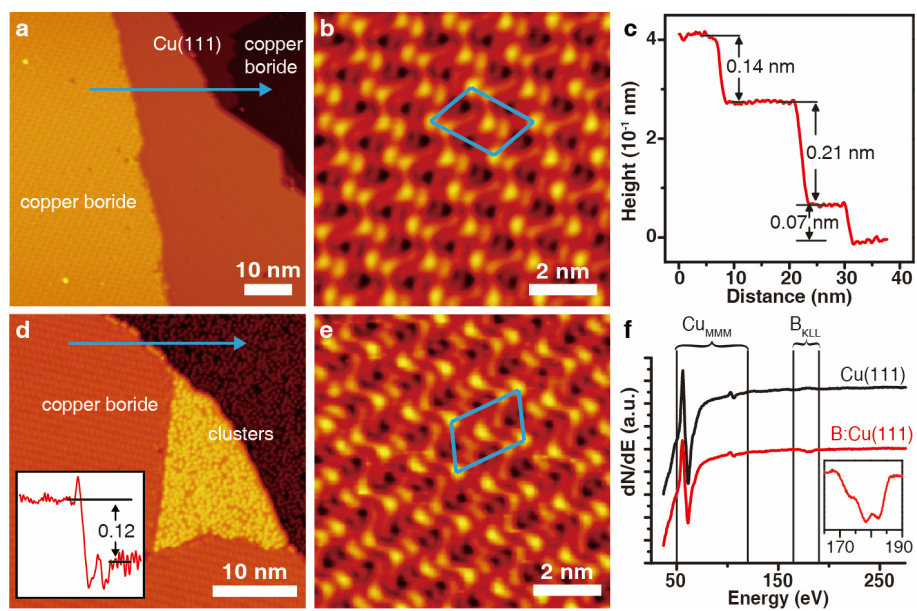

**Fig. 1 Characterization of copper boride on Cu(111). a**, STM image of the surface after the boron deposition at room temperature and subsequent annealing at 540 K ($V_t$ = 1.00 V, $I_t$ = 90 pA). **b**, High-resolution STM image of copper boride in **a** ($V_t$ = 0.97 V, $I_t$ = 220 pA). **c**, Profile of the cross section indicated by the arrow in **a**. **d**, STM image of the surface after boron deposition at 740 K ($V_t$ = -1.90 V, $I_t$ = 90 pA). The inset shows the line profile of the cyan arrow. **e**, High-resolution STM image of copper boride in **d** ($V_t$ = 1.44 V, $I_t$ = 150 pA). **f**, AES spectra of Cu(111) before and after boron deposition.

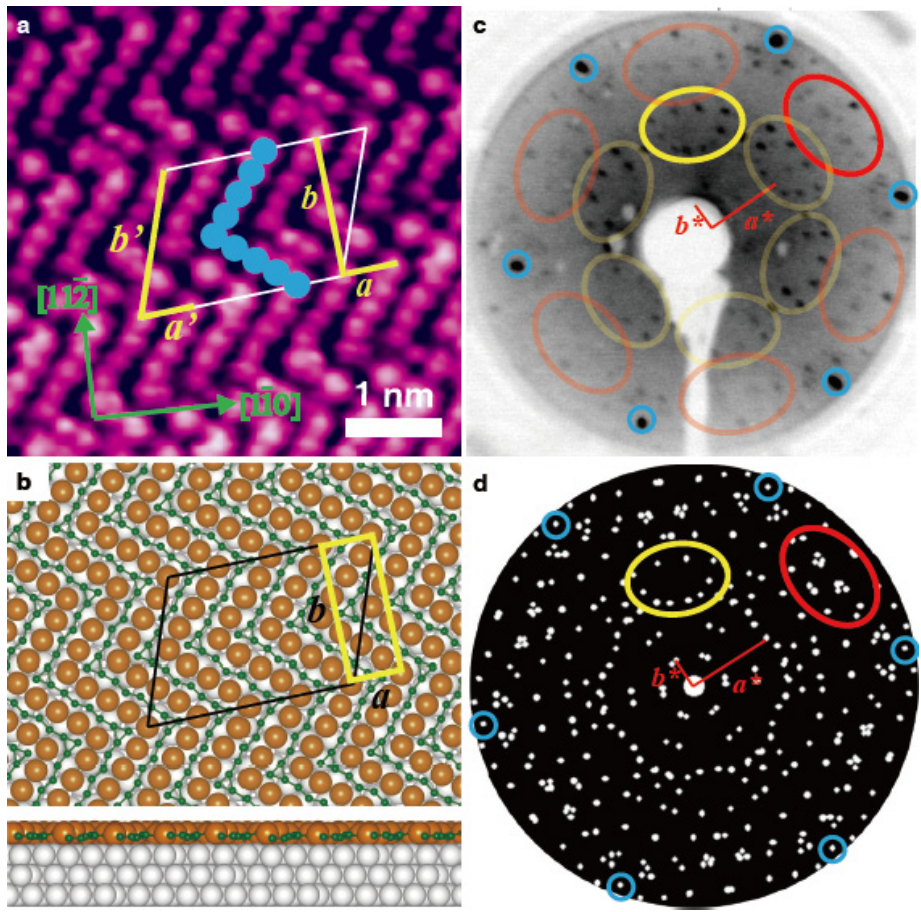

**Fig. 2 Lattice vectors of copper boride. a**, Atomic-resolution STM image of copper boride ($V_t$ = -0.033 V, $I_t$ = 1570 pA). Two equivalent lattice vectors (*a, b* and *a', b'*) for copper boride are marked by yellow lines. The white frame corresponds to the $\sqrt{73} \times \sqrt{39}$ superstructure of Cu(111). **b**, Atomic configuration of copper boride on Cu(111). Orange, green and white spheres represent Cu and B atoms of copper boride and Cu atoms of the substrate. The unit cells with and without the substrate are colored in black and yellow, respectively. **c**, Experimental LEED patterns for the sample with fully grown copper boride on Cu(111) with the energy of 55.8 eV. **d**, Simulated LEED patterns of Cu(111) overlapped by copper boride. Reciprocal lattice vectors ($a^*$, $b^*$) are marked in red. Some characteristic spots are marked by blue circles, red and yellow ovals to help compare **c** and **d**.

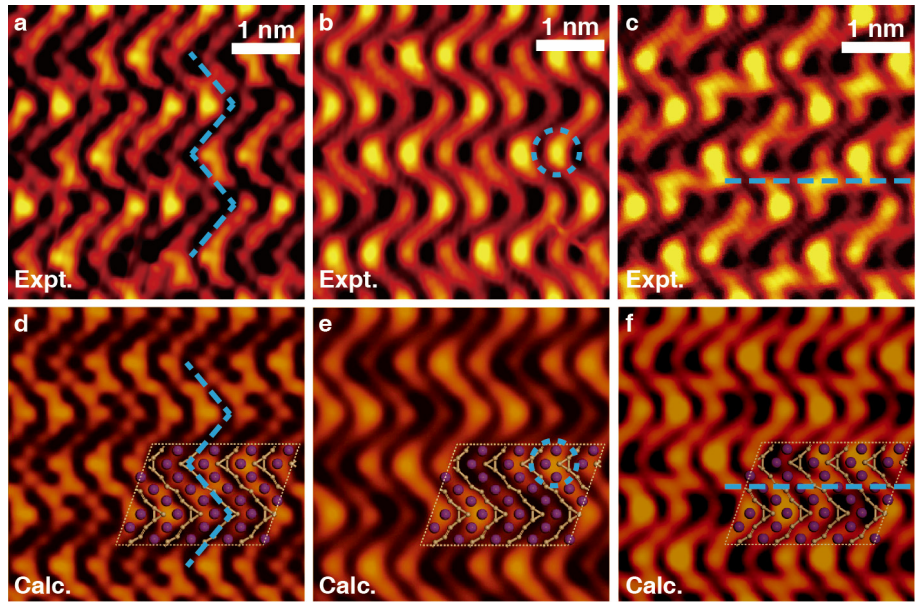

**Fig. 3 Comparison of experimental results and simulated STM images for 2D-$Cu_8B_{14}$ on Cu(111).** Experimental STM images taken at **a,** $V_t$ = 0.106 V, $I_t$ = 510 pA, **b**, $V_t$ = 0.311 V, $I_t$ = 340 pA, **c**, $V_t$ = -1.632 V, $I_t$ = 790 pA. Constant current STM simulations were performed at the biases of **d**, 1.0, **e**, 1.5, and **f**, -1.6 eV. The blue zigzag lines, circles and dotted lines with the overlaid atomic structures are provided for reference.

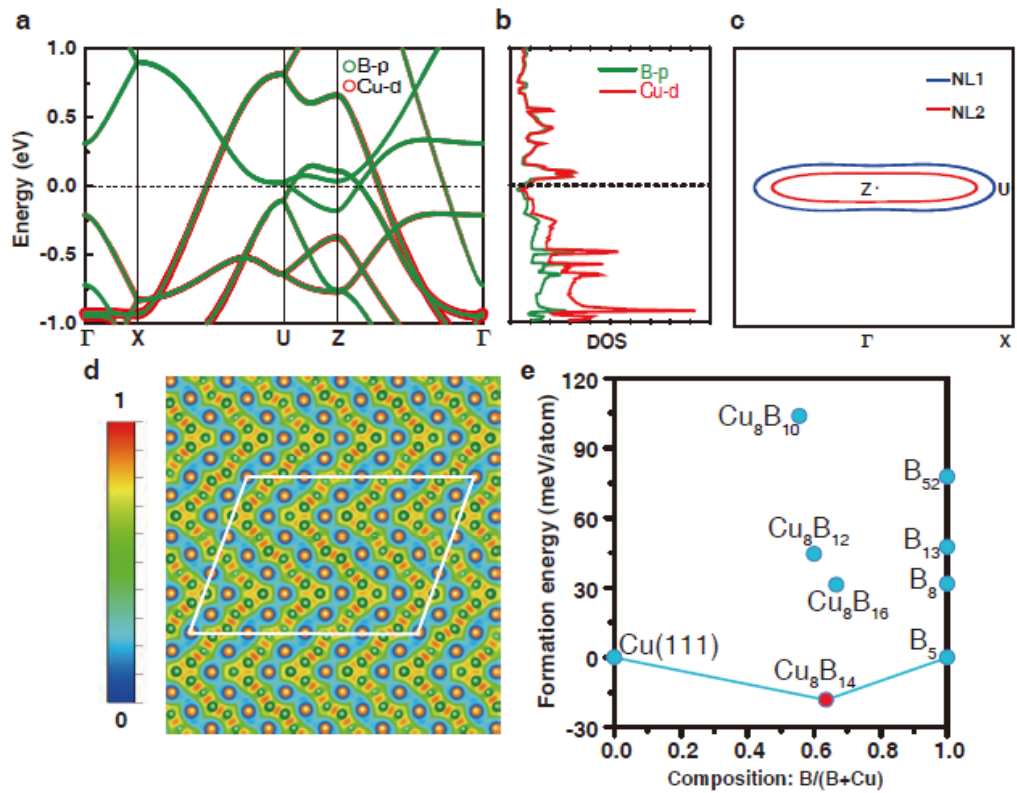

**Fig. 4 Electronic properties and thermodynamic stability of $Cu_8B_{14}$. a**, Orbital-projected band structure of freestanding $Cu_8B_{14}$. **b**, Projected density of states of freestanding $Cu_8B_{14}$. **c,** Nodal lines (NL1 and NL2) around the *Z* point. **d**, ELF of $Cu_8B_{14}$ on Cu(111). Cu and B atoms are colored in orange and green, respectively. **e**, Convex hull for a variety of related structures on Cu(111).